\documentclass[12pt]{article}
\usepackage{amsmath}
\usepackage{latexsym}
\usepackage[dvips]{graphicx}
\usepackage{epsfig}
\usepackage[hang]{caption2}

\usepackage{newcent}

\begin{document}

\begin{center}
{\Large Arbitrage Opportunities and their Implications to Derivative
Hedging}

{\Large} \bigskip

Stephanos Panayides \vskip 1.0cm

The School of Mathematics, The University of Manchester

Manchester, M60 1QD, UK

Tel: +44 (0) 161 200 3696

E-mail: stephanos.panayides@student.manchester.ac.uk

\bigskip

Submitted to Physica A on 7 February 2005

\bigskip

\bigskip\
\textbf{Abstract}
\end{center}

We explore the role that random arbitrage opportunities play in
hedging financial derivatives. We extend the asymptotic pricing
theory presented by Fedotov and Panayides [Stochastic arbitrage
return and its implication for option pricing, Physica A 345 (2005),
207-217] for the case of hedging a derivative when arbitrage
opportunities are present in the market. We restrict ourselves to
finding hedging confidence intervals that can be adapted to the
amount of arbitrage risk an investor will permit to be exposed to.
The resulting hedging bands are independent of the detailed
statistical characteristics of the arbitrage opportunities.

\bigskip
\noindent PACS number: 02.50.Ey; 89.65.Gh

\begin{description}
\item[Keywords:] Derivative hedging, random arbitrage, hedging ratio.
\end{description}

\pagebreak

\section{Introduction}

In the classical Black-Scholes methodology a perfect hedge is
achieved by holding an amount $\frac{\partial V}{\partial S}$ of the
underlying asset $S$ in order to eliminate the risk from the changes
in the stock price \cite{14}. When one of the Black-Scholes ideal
assumptions is relaxed the procedure is less clear and alternative
methods have to be proposed (see for example \cite{15,16,17}). In
this paper we relax the no-arbitrage assumption and study the
hedging problem for the case arbitrage opportunities are stochastic.

Empirical studies have shown existence of short lived arbitrage
opportunities \cite{1,2}. Thus, accepting that arbitrage
opportunities do exist, one is forced to consider them when pricing
and hedging a derivative. The first attempt to take into account
arbitrage opportunities for pricing a derivative is given in
\cite{3,4} where the constant interest rate $r_{0}$ is substituted
by the stochastic process $r_{0}+x(t)$. The random arbitrage $x(t)$
is assumed to follow an Ornstein-Uhlenbeck process. In \cite{5,10}
the same idea is reformulated in terms of option pricing with
stochastic interest rate. For a more recent literature on arbitrage
we refer the reader to \cite{20,30} and references therein.

In \cite{6} Fedotov and Panayides present an asymptotic pricing
theory, based on the central limit theorem \cite{8}, for pricing an
option contract. The methodology focuses on deriving pricing bands
for the option rather than finding an exact equation for the price.
The nice feature of the method is that the pricing bands are
independent of the detailed statistical characteristics of the
random arbitrage. In this paper we apply the same methodology as for
the pricing case and try to find hedging confidence intervals that
can be adapted to the amount of arbitrage risk an investor will
permit to be exposed to.

In section 2 we briefly review the methodology used in \cite{6}. In
section 3 we use the same technique to find hedging bands around the
usual Black-Scholes hedging ratio that account for the arbitrage
opportunities. Numerical calculations are presented in section 4.
Concluding remarks are given in section 5.

\section{Methodology}

Following \cite{6}, consider a market that consists of a stock $S$,
a bond $B$, and a European option on the stock $V$. The market is
assumed to be affected by two sources of uncertainty, the random
fluctuations of the return from the stock, and a random arbitrage
return from the bond described respectively by the conventional
stochastic differential equations
\begin{equation}
\frac{dS}{S}=\mu dt+\sigma dW,  \label{stock}
\end{equation}
and
\begin{equation}
\frac{dB}{B}=rdt+\xi (t)dt,  \label{bond}
\end{equation}
where $r$ is the risk-free interest rate, and $W$ the standard
Wiener process. The random process $\xi (t)$  describes the
fluctuations of the arbitrage return around $rdt$.

The random variations of arbitrage return $\xi
(t)$ are assumed to be on the scale of hours. Denote this characteristic time by $%
\tau _{arb}$ and regard it as an intermediate one between the time
scale of random stock return (infinitely fast Brownian motion
fluctuations), and the lifetime of the derivative $T$ (several
months): $0<<$ $\tau _{arb}<<T$. This difference in time allows the
development of an asymptotic pricing theory involving the central
limit theorem for random processes.

The next step is the derivation of the PDE satisfying the option
price $V$. Consider the investor establishing a zero initial
investment position by creating
a portfolio $\Pi $ consisting of one bond $B,$ $-\frac{%
\partial V}{\partial S}$ shares of the stock $S$ and one European option $V$ with an exercise price $K$ and a maturity
$T.$ The value of this portfolio is given by
\begin{equation}
\Pi =B-\frac{\partial V}{\partial S}S+V.  \label{portfolio}
\end{equation}
The Black-Scholes dynamic of this portfolio is given by the two equations $%
\partial \Pi /\partial t=0$ and $\ \Pi =0$ . The
application of It{\^o}'s formula to (\ref{portfolio}), together with (\ref{stock}%
) and (\ref{bond}) with $\xi (t)=0,$ leads to the classical
Black-Scholes equation:
\begin{equation}
\frac{\partial V}{\partial t}+\frac{\sigma ^{2}S^{2}}{2}\frac{\partial ^{2}V%
}{\partial S^{2}}+rS\frac{\partial V}{\partial S}-rV=0.  \label{BS}
\end{equation}
A generalization of $\partial \Pi /\partial t=0$ is the simple
non-equilibrium equation
\begin{equation}
\frac{\partial \Pi }{\partial t}=-\frac{\Pi }{\tau _{arb}},
\label{p}
\end{equation}
where $\tau _{arb}$ is the characteristic time during which the
arbitrage opportunity ceases to exist (see \cite{9}). Using the self-financing condition $d\Pi =dB-\frac{\partial V}{\partial S}%
dS+dV$, Ito's lemma, and equations (\ref{stock}) and (\ref{bond}),
the equation for the option value $V(S,t)$ is easily shown to
satisfy the PDE
\begin{equation}
\frac{\partial V}{\partial t}+\frac{\sigma ^{2}S^{2}}{2}\frac{\partial ^{2}V%
}{\partial S^{2}}+rS\frac{\partial V}{\partial S}-rV+r\Pi +\xi
(t)\Pi +\xi
(t)\left( S\frac{\partial V}{\partial S}-V\right) +\frac{\Pi }{\tau _{arb}}%
=0.  \label{basic}
\end{equation}
This equation reduces to the classical Black-Scholes PDE (\ref{BS})
when $\Pi =0$ and $\xi (t)=0.$

To deal with the forward problem lets introduce the non-dimensional
time
\begin{equation}
\tau =\frac{T-t}{T},\;\;\;0\leq \tau \leq 1,  \label{time}
\end{equation}
and a small parameter
\begin{equation}
\varepsilon =\frac{\tau _{arb}}{T}<<1.
\end{equation}
In the limit $%
\varepsilon \rightarrow 0,$ the stochastic arbitrage return $\xi $
becomes a function that is rapidly varying in time, say $\xi \left(
\frac{\tau }{\varepsilon }\right) $ (see \cite{44}). The random
arbitrage $\xi(\tau)$ is assumed to be an ergodic random process
with
\ $<$ $%
\xi (\tau )$ $>=0,$ such that
\begin{equation}
D=\int_{0}^{\infty }<\xi (\tau +s)\xi (\tau )>ds  \label{dif}
\end{equation}
is finite.
From (\ref{p}) the value of the portfolio $%
\Pi$ decreases to zero exponentially. Thus, it is assumed that $\Pi
$ $=0$ in the limit $\varepsilon \rightarrow 0$. From (\ref{basic})
the associated option price $V^{\varepsilon}=V^{\varepsilon }\left(
\tau ,S\right) $ obeys the stochastic PDE
\begin{equation}
\frac{\partial V^{\varepsilon }}{\partial \tau }=\frac{\sigma ^{2}S^{2}}{2}%
\frac{\partial ^{2}V^{\varepsilon }}{\partial
S^{2}}+rS\frac{\partial
V^{\varepsilon }}{\partial S}-rV^{\varepsilon }+\xi (\frac{\tau }{%
\varepsilon })\left( S\frac{\partial V^{\varepsilon }}{\partial S}%
-V^{\varepsilon }\right),   \label{basic2}
\end{equation}
with the initial condition
\begin{equation}
V^{\varepsilon }(S,0)=\max (S-K,0)
\end{equation}
for a call option with strike price $K$. Using the law of large
numbers and the central limit theorem for stochastic processes (see
\cite{8}), Fedotov and Panayides analyze the stochastic PDE
(\ref{basic2}) and give pricing bands for the option. More details
are given in \cite{6}.

\section{Hedging Ratio}

In this section we address the problem of hedging the risk from
writing an option on an asset, for the case stochastic arbitrage
opportunities are present in the market. Here, we apply the same
methodology as for the pricing case used in \cite{6}, and try to
find hedging confidence intervals that can be adapted to the amount
of arbitrage risk an investor will permit to be exposed to. The
method can be used to save on the cost of hedging in a random
arbitrage environment.

In the usual Black-Scholes case a perfect hedge is achieved by
holding an amount $\Delta_{BS} =%
\frac{\partial V}{\partial S}$ of the underlying asset in order to
eliminate the risk from the changes in the stock price.
Differentiating (\ref{BS}) with respect to $S$, and introducing the
non-dimensional time $\tau$ as before, gives
\begin{equation}
\frac{\partial \Delta_{BS}}{\partial \tau }=S(\sigma
^{2}+r)\frac{\partial \Delta_{BS}}{\partial S}+\frac{\sigma
^{2}S^{2}}{2}\frac{\partial ^{2}\Delta_{BS}}{\partial S^{2}},
\label{BSHR}
\end{equation}
with initial condition $\Delta _{BS}\left( S,0\right) =H(S-K)$ for a
call option with strike price $K$ and $H$ the Heaviside
function\footnote{The Heaviside function $H(x)$ is defined by
$H(x)=0$ if $x<0$ and $H(x)=1$ if $x>0$.}. Similarly,
differentiating (\ref{basic2}) with respect to $S$ the stochastic
hedging ratio that accounts for the stochastic arbitrage returns,
defined by $\Delta ^{\varepsilon }=\frac{\partial V^{\varepsilon
}}{\partial S}$ with $\varepsilon $ as before, satisfies the PDE
\begin{equation}
\frac{\partial \Delta ^{\varepsilon }}{\partial \tau }=\sigma ^{2}S\frac{%
\partial \Delta ^{\varepsilon }}{\partial S}+\frac{\sigma ^{2}S^{2}}{2}\frac{%
\partial ^{2}\Delta ^{\varepsilon }}{\partial S^{2}}+(r+\xi (\frac{\tau }{%
\varepsilon }))S\frac{\partial \Delta ^{\varepsilon }}{\partial S}.
\label{HPDE}
\end{equation}
According to the law of large numbers, the stochastic hedging ratio
$\Delta^{ \varepsilon }=\Delta ^{\varepsilon }\left( S,\tau\right)$
converges in probability to the Black-Scholes hedging ratio $\Delta
_{BS}\left( S,\tau\right)$. Splitting $\Delta ^{\varepsilon }=%
\frac{\partial V^{\varepsilon }}{\partial S},$ into the sum of the
Black-Scholes hedging ratio $\Delta _{BS}\left( S,\tau\right) ,$ and
the random field $Y^{\varepsilon }\left( S,\tau\right) ,$ with the
scaling factor $\sqrt{\varepsilon }$, gives
\begin{equation}
\Delta ^{\varepsilon }\left( S,\tau \right) =\Delta _{BS}\left(
S,\tau\right) +\sqrt{\varepsilon }Y^{\varepsilon }\left(
S,\tau\right) . \label{bands}
\end{equation}
Substituting (\ref{bands}) into (\ref{HPDE}), and using equation
(\ref{BSHR}), we get the stochastic PDE for the random field
$Y^{\varepsilon }\left( S ,\tau\right)$, given by
\begin{equation}
\frac{\partial Y^{\varepsilon }}{\partial \tau }=\frac{1}{2}\sigma ^{2}S^{2}%
\frac{\partial ^{2}Y^{\varepsilon }}{\partial S^{2}}+(\sigma ^{2}S+rS+\xi (%
\frac{\tau }{\varepsilon })S)\frac{\partial Y^{\varepsilon }}{\partial S}%
+\xi (\frac{\tau }{\varepsilon })\frac{S}{\sqrt{\varepsilon
}}\frac{\partial \Delta _{BS}}{\partial S}.
\end{equation}
In what follows, we try to find the asymptotic equation for
$Y^{\varepsilon }\left( S,\tau\right) $ as $\varepsilon \rightarrow
0.$ Applying Ergodic theory, in the limit $\varepsilon \rightarrow
0,$ the random field $Y^{\varepsilon }\left( S,\tau\right)$
converges weakly to the field $Y\left( S,\tau \right)$ satisfying
the linear stochastic PDE
\begin{equation}
\frac{\partial Y}{\partial \tau }=\frac{1}{2}\sigma
^{2}S^{2}\frac{\partial ^{2}Y}{\partial
S^{2}}+S\left( \sigma ^{2}+r\right) \frac{\partial Y}{\partial S}+S\frac{%
\partial \Delta _{BS}}{\partial S}\eta \left( \tau \right) ,  \label{s1}
\end{equation}
for $\eta \left( \tau \right) $ the white Gaussian noise satisfying
\begin{equation}
<\eta (\tau _{1})\eta (\tau _{2})>=2D\delta \left( \tau _{1}-\tau
_{2}\right) ,  \label{delta}
\end{equation}
$D$ the diffusion constant given by (\ref{dif}), and with initial
condition $Y\left( S,0\right) =0.$ The solution to equation
(\ref{s1}) is given by
\begin{equation}
Y(\tau ,S)=\int_{0}^{\tau }\int_{0}^{\infty }G(S,S_{1},\tau ,\tau _{1})S_{1}%
\frac{\partial \Delta _{BS}}{\partial S_{1}}\eta \left( \tau
_{1}\right) dS_{1}d\tau _{1},  \label{wc}
\end{equation}
where $G(S,S_{1},\tau ,\tau _{1})$ is the Green's function
corresponding to (\ref{s1}) (see \cite{11}). From the solution
(\ref{wc}) it follows that \ $Y(S,\tau)$ is a Gaussian field with
zero mean and covariance $R(S,x,\tau)=<Y\left( S,\tau\right)
Y(x,\tau)>$ satisfying the deterministic PDE
\begin{multline}
\frac{\partial R}{\partial \tau }=\frac{1}{2}\sigma
^{2}S^{2}\frac{\partial
^{2}R}{\partial S^{2}}+\frac{1}{2}\sigma ^{2}x^{2}\frac{\partial ^{2}R}{%
\partial x^{2}}+S\left( \sigma ^{2}+r\right) \frac{\partial R}{\partial S}%
+x\left( \sigma ^{2}+r\right) \frac{\partial R}{\partial x}+\\2D\left( S\frac{%
\partial \Delta _{BS}}{\partial S} \times x\frac{\partial \Delta _{BS}}{\partial x}%
\right) ,  \label{stef2}
\end{multline}
with $R(S,x,0)\equiv 0$. The typical hedging bands for the case of
arbitrage opportunities can be given by
\begin{equation}
\Delta _{BS}\left( S,\tau\right) \pm 2\sqrt{\varepsilon U\left(
S,\tau\right) }.
\end{equation}
Hedging within two standard deviations of the Black-Scholes hedging
ratio provides a higher protection against arbitrage fluctuations.
The variance $U\left( S,\tau\right)=R(S,S,\tau)$ quantifies the
fluctuations around the Black-Scholes hedging ratio and is given by
\begin{equation}
U\left( S,\tau\right)=2D\int_{0}^{\tau }[\int_{0}^{\infty
}G\left( S,S_{1},\tau ,\tau _{1}\right) S_{1}\frac{\partial \Delta _{BS}}{%
\partial S_{1}}dS_{1}]^{2}d\tau _{1}.\label{rim}
\end{equation}
One can conclude that the investor hedges the option using the
hedging ratio
\begin{equation}
\Delta _{BS}\left( S,\tau\right) + 2\sqrt{\varepsilon U\left(
S,\tau\right) }.\label{kotsirouin}
\end{equation}

\section{Numerical Results}

From equation (\ref{s1}) or (\ref{rim}) we can see that the large fluctuations of $Y(S,\tau)$ occur when the function $S\frac{%
\partial \Delta _{BS}}{\partial S}$ takes its maximum value. This is the case for near at-the money options.
Thus, the risk error (bandwidth) in hedging an option is greater for
these cases. Using equation (\ref{stef2}) we get a plot of the
covariance $U(S,\tau)$ with respect to asset price $S$ and time
$\protect\tau$ (Figure 1). From the graph we observe that the
hedging error increase as we move near at-the-money. This result is
consistent with the empirical results presented in \cite{12,13}. In
Figure 2, we plot the effective hedging ratio given by equation
(\ref{kotsirouin}), for $\varepsilon=0.1$, and compare it with the
usual Black-Scholes hedging ratio. Note that the largest deviations
from the usual Black-Scholes hedging ratio are near at-the money,
while they decrease as we move in and out-of-the-money. In
particular, the arbitrage corrections for near at-the-money account
for approximately 3.5\% change in the Black-Scholes hedging ratio.

\section{Conclusions}

Using an asymptotic pricing theory, first introduced by Fedotov and
Panayides \cite{6}, we explored the role that random arbitrage
opportunities play in hedging derivatives. In particular, we managed
to give hedging bands around the usual Black-Scholes hedging ratio
that account for the stochastic nature of arbitrage opportunities.
Numerical results showed that the largest deviations from the usual
Black-Scholes hedging ratio are near at-the money. Note that the
work in this paper is purely theoretical. Despite this the results
are consistent with empirical work in the literature. In future work
we plan to address parametric and non-parametric statistical tests
on a large sample of observations of trades to explain
quantitatively any market deviations from the Black-Scholes price
and hedge ratio for the case of arbitrage returns.

\pagebreak

\linespread{0.4}
\begin{figure}[p]
\centering
\includegraphics[scale=0.7]{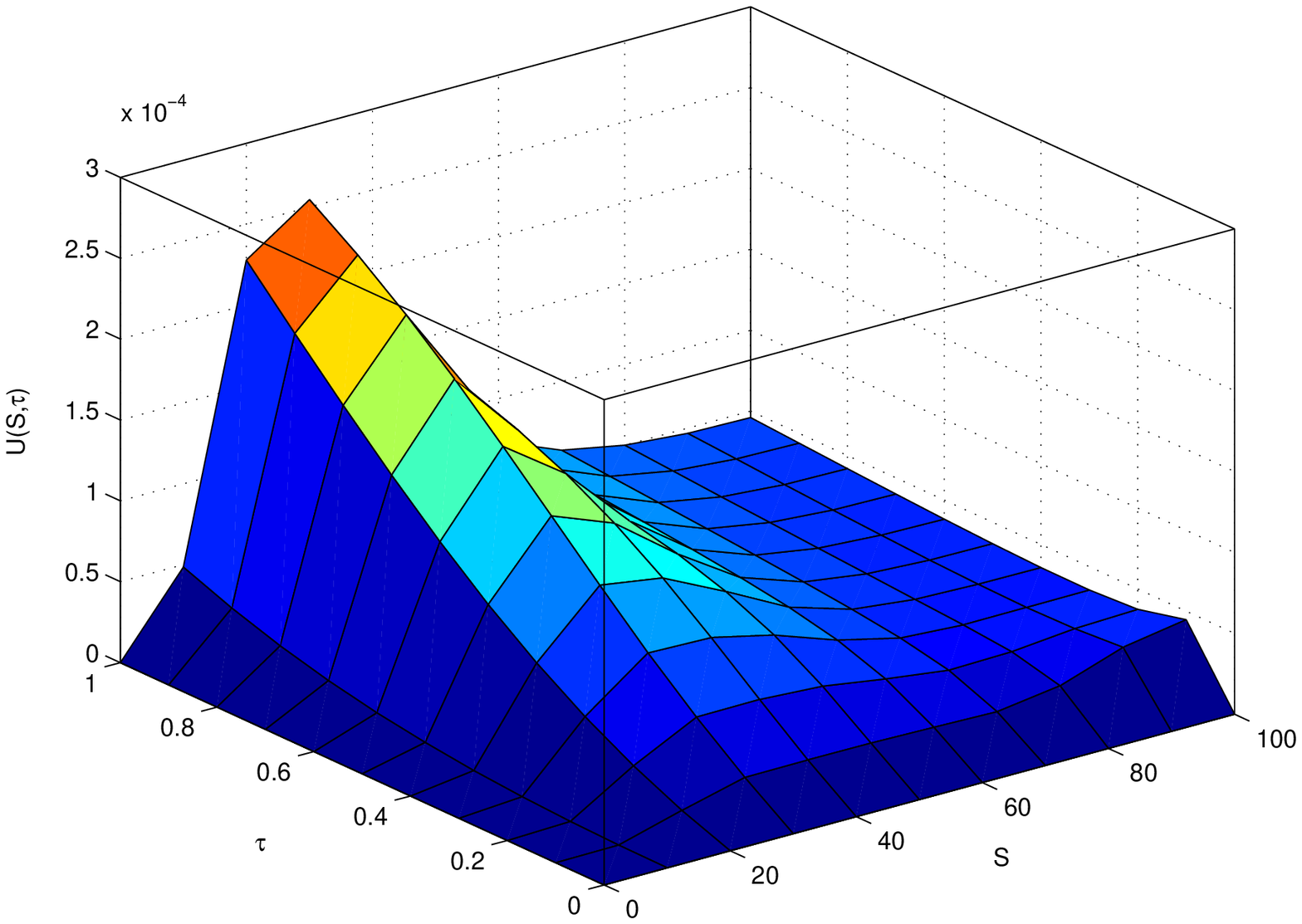}
\caption[Hedging fluctuations around Black-Scholes hedging ratio,
due to arbitrage returns.]{Variance $U(S,\protect\tau)$, with
respect to asset price $S$, and time $\protect\tau $. The option
strike price $K=20$, the volatility $\protect\sigma=0.4$, the
interest rate $r=0.1$, and the constant $D=0.1$.}
\end{figure}

\linespread{0.4}
\begin{figure}[p]
\centering
\includegraphics[scale=1.0]{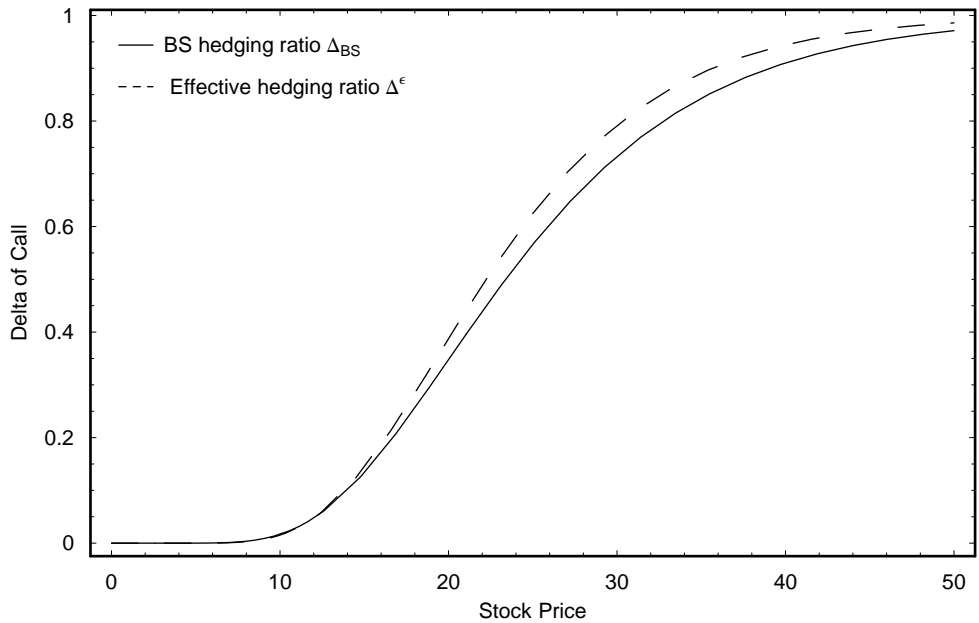}
\caption[Comparison of effective hedging ratio for arbitrage
returns, and Black-Scholes hedging ratio.]{Effective hedging ratio
and Black-Scholes hedging ratio. Here $\varepsilon=0.1$, $T=1$
(years) and constants $K$, $\sigma$, $r$, and $D$ are taken as in
Figure 1.}
\end{figure}

\end{document}